\documentclass[apl,amsmath,reprint,floatfix,a4paper,superscriptaddress]{revtex4-1}

\usepackage[T1]{fontenc}
\usepackage{etex}
\usepackage{geometry}
\usepackage{color,tabularx,colortbl}

\usepackage{setspace}
\usepackage{pgfplots}
\usepackage{booktabs}
\usepackage{multirow}
\usepackage{tikz, tikz-3dplot}
\tikzset{>=latex}
\usepackage[percent]{overpic}
\pgfplotsset{compat=1.13}
\usepgfplotslibrary{external} 
\usepgfplotslibrary{fillbetween}
\usepgfplotslibrary{colorbrewer}
\usetikzlibrary{pgfplots.groupplots,arrows,decorations.markings}
\usetikzlibrary{shapes.geometric,calc,patterns}
\usepackage[strict]{changepage}

\definecolor{uni_blue}{RGB}{0,99,166}
\definecolor{uni_grey}{RGB}{102,102,102}
\definecolor{uni_red}{RGB}{167,28,73}
\definecolor{uni_orange}{RGB}{221,72,20}
\definecolor{uni_yellow}{RGB}{246,168,0}
\definecolor{uni_light_green}{RGB}{148,193,84}
\definecolor{uni_mint}{RGB}{17,137,122}

\pgfplotsset{
colormap={myColMap}{[5pt] 
	    color(0pt)=(white);
        color(10pt)=(uni_blue!10);
        color(20pt)=(uni_blue!20);
        color(30pt)=(uni_blue!30);
        color(40pt)=(uni_blue!40);
        color(50pt)=(uni_blue!50);
        color(60pt)=(uni_blue!60);
        color(70pt)=(uni_blue!70);
        color(80pt)=(uni_blue!80);
        color(90pt)=(uni_blue!90);
        color(100pt)=(uni_blue);
	    }
}
\pgfplotsset{
colormap={myColMapReduced}{[5pt] 
	    color(0pt)=(white);
	    color(500pt)=(white);
	    color(505pt)=(uni_blue);
        color(1000pt)=(uni_blue);
	    }
}

\pgfplotscreateplotcyclelist{uni}{
    {color=uni_blue,mark=none},
    {dashed,color=uni_yellow,mark=none},
    {color=uni_red,mark=none},
    {dashed,color=uni_light_green,mark=none},
    {color=uni_orange,mark=none},
    {dashed,color=uni_grey,mark=none},
    {color=uni_light_green,mark=none},
    }

\begin{document}

\title{Hysteresis-free magnetization reversal of exchange-coupled bilayers with finite magnetic anisotropy} 

\author{Christoph Vogler}
\email{christoph.vogler@univie.ac.at}
\affiliation{Faculty of Physics, University of Vienna, Boltzmanngasse 5, 1090 Vienna, Austria}
\author{Michael Heigl}
\affiliation{Institute of Physics, University of Augsburg, Augsburg 86159, Germany}
\author{Andrada-Oana Mandru}
\affiliation{Empa--Swiss Federal Laboratories for Materials Science and Technology, CH-8600 D{\"u}bendorf, Switzerland}
\author{Birgit Hebler}
\affiliation{Institute of Physics, University of Augsburg, Augsburg 86159, Germany}
\author{Miguel Marioni}
\affiliation{Empa--Swiss Federal Laboratories for Materials Science and Technology, CH-8600 D{\"u}bendorf, Switzerland}
\author{Hans Josef Hug}
\affiliation{Empa--Swiss Federal Laboratories for Materials Science and Technology, CH-8600 D{\"u}bendorf, Switzerland}
\affiliation{Department of Physics, University of Basel, CH-4056 Basel, Switzerland}
\author{Manfred Albrecht}
\affiliation{Institute of Physics, University of Augsburg, Augsburg 86159, Germany}
\author{Dieter Suess}
\affiliation{Christian Doppler Laboratory for Advanced Magnetic Sensing and Materials, Faculty of Physics, University of Vienna, Boltzmanngasse 5, 1090 Vienna, Austria}

\begin{abstract}
Exchange-coupled structures consisting of ferromagnetic and ferrimagnetic layers become technologically more and more important. We show experimentally the occurrence of completely reversible, hysteresis-free minor loops of [Co(0.2\,nm)/Ni(0.4\,nm)/Pt(0.6\,nm)]$_N$ multilayers exchange-coupled to a 20\,nm thick ferrimagnetic Tb$_{28}$Co$_{14}$Fe$_{58}$ layer, acting as hard magnetic pinning layer. Furthermore, we present detailed theoretical investigations by means of micromagnetic simulations and most important a purely analytical derivation for the condition of the occurrence of full reversibility in magnetization reversal. Hysteresis-free loops always occur if a domain wall is formed during the reversal of the ferromagnetic layer and generates an intrinsic hard-axis bias field that overcomes the magnetic anisotropy field of the ferromagnetic layer. The derived condition further reveals that the magnetic anisotropy and the bulk exchange of both layers, as well as the exchange coupling strength and the thickness of the ferromagnetic layer play an important role for its reversibility.
\end{abstract}
\maketitle

\section{Introduction}
\pgfplotsset{colormap/Blues-9}
Ferrimagnets (FI) are becoming technologically more and more important to replace antiferromagnets when it is required to pin the magnetization of a ferromagnet (FM) to a certain direction~\cite{romer_temperature_2012}. Typically the magnetization process of a ferromagnetic thin film with strong magnetic anisotropy and a magnetic field applied along the easy axis of magnetization is expected to be hysteretic. The size of the reversal field depends on domain wall pinning but always remains smaller than the anisotropy field. Ferromagnetic films coupled to antiferromagnets typically show exchange bias behavior, i.e. the hysteresis loop is shifted by the so-called exchange field in the horizontal direction but can also be shifted in the vertical direction. The latter is usually attributed to pinned uncompensated spins, with a part of them responsible for the exchange bias effect~\cite{schmid_role_2007}. Typically the width of the hysteresis loop is increased that is often attributed to the coupling of the ferromagnetic moments to rotating uncompensated spins, but was also shown to arise from the motion of ferromagnetic domains over the inhomogeneous spatial distribution of interfacial pinned uncompensated spin density~\cite{benassi_role_2014}. In contrast to antiferromagnets, ferrimagnets offer a high degree of design flexibility. Antiferromagnetically exchange coupled ferro-/ferrimagnetic bilayers have been investigated by Mangin et al.~\cite{mangin_influence_2008}. In their work they identified the magnetic wall configuration at the interface as the determining mechanism for the exchange bias field. Further systematic studies were performed to investigate the impact of the Fe-Co ratio on the exchange coupling in TbFeCo/[Co/Pt] heterostructures as well as  their dependence on the composition of the ferrimagnetic layer and number of repetitions of the [Co/Pt] multilayer~\cite{hebler_influence_2016,oezelt_micromagnetic_2015,schubert_interfacial_2013}. Of particular interest are ferromagnetic layers that are exchange-coupled to a highly coercive ferrimagnetic pinning layer. In this case, a giant exchange field occurs that remains however limited by the coercive field of the pinning layer~\cite{becker_switching-field_1997,mangin_interface_2003,lin_high_2003,watson_interfacial_2008,romer_temperature_2012,schubert_interfacial_2013,hebler_influence_2016}.

In this regard, very recently the complex magnetization reversal of a ferromagnetic [Co(0.4\,nm)/Pt(0.7\,nm)]$_5$ multilayer exchange-coupled to a ferrimagnetic Tb$_{26.5}$Co$_{73.5}$ film was investigated~\cite{zhao_magnetization_2019}. The complex reversal process was attributed to the spatial variation of the material properties of the ferromagnetic [Co/Pt] multilayer deposited on top of the hard magnetic Tb$_{26.5}$Co$_{73.5}$ pinning layer. Micromagnetic simulations have revealed that [Co/Pt] grains with hysteretic and non-hysteretic reversal can coexist but will still lead to an (almost) hysteresis-free reversal of the macroscopic magnetization loop. Being able to design a system with a hysteresis-free magnetization process would offer additional features, providing great potential for many applications, i.e. magnetic sensors.

The magnetization reversal in thin ferromagnetic films is often analyzed in the context of the Stoner-Wohlfarth model. However, in this model the hysteresis width  for a field applied along the easy axis increases linearly with increasing magnetic anisotropy and is only zero for vanishing anisotropy. To the best of our knowledge, there exists neither a physical explanation nor an analytical expression for the occurrence of hysteresis-free loops of thin ferromagnetic layers with finite magnetic anisotropy. In this work, we show a systematic experimental and theoretical study of this phenomena and reveal the physical conditions of hysteresis-free magnetization reversal.

\section{Experiments}
\label{sec:exp}
All samples were prepared on Si(001)/SiO$_2$(100\,nm) substrates at room temperature using magnetron (co-)sputtering from elemental targets. For the depositions the Ar pressure was kept constant at $5 \times 10^{-3}$\,mbar and the base pressure remained below $1 \times 10^{-8}$\,mbar. The exchange-coupled heterostructures consist of a 20\,nm-thick amorphous ferrimagnetic Tb$_{28}$Co$_{14}$Fe$_{58}$ layer with a ferromagnetic [Co(0.2\,nm)/Ni(0.4\,nm)/Pt(0.6\,nm)]$_N$ multilayer on top. The number of repetitions, $N$ was varied and with it the total thickness of the ferromagnetic layer. 
In addition to these exchange-coupled FM/FI samples, reference samples consisting of only the ferromagnetic or ferrimagnetic layers were fabricated to extract their magnetization and magnetic anisotropy as input for the micromagnetic calculations. 
For all samples, 5\,nm of Pt were used both as seed layer and as a capping layer. 
\begin{figure}
    \centering
    \includegraphics[width=0.9\linewidth]{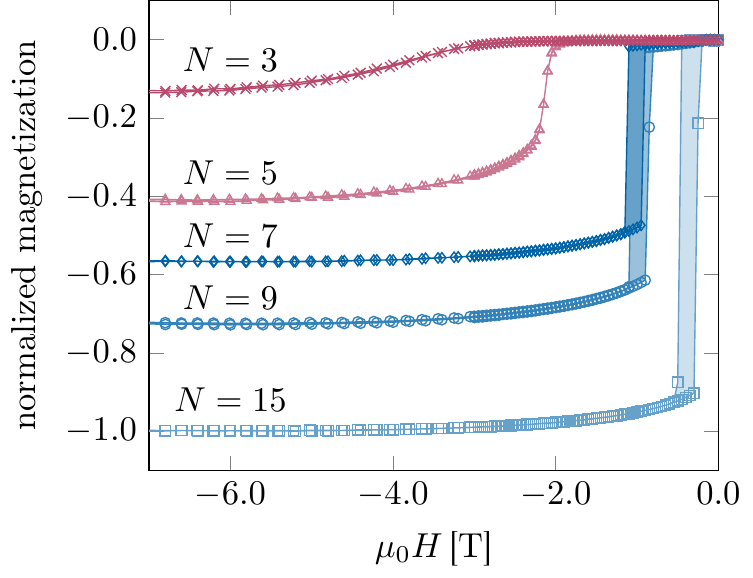}
    \caption{\small (color online) Easy-axis minor loops of ferromagnetic [Co(0.2\,nm)/Ni(0.4\,nm)/Pt(0.6\,nm)]$_N$ multilayers antiferromagnetically coupled to a 20\,nm thick TbCoFe layer at 40\,K for various repetition numbers $N$ of the ferromagnetic layer.}
    \label{fig:exp}
\end{figure}
$M$-$H$ hysteresis loops were acquired with a superconducting quantum interference device-vibrating sample magnetometer (SQUID-VSM). All measurements were performed at 40\,K with a maximum external field of 7\,T. In order to determine the magnetic properties, SQUID-VSM measurements in both out-of-plane (oop) and in-plane (ip) geometries were performed. The effective magnetic anisotropies $K_{\mathrm{eff}}$ were extracted from the differences of the areas enclosed by the hard axis (ip) and averaged easy-axis (oop) loops and the magnetization axis of the $M$-$H$ loops. 
All [Co/Ni/Pt] multilayer films show a strong perpendicular magnetic anisotropy (PMA) with  $K_\mathrm{eff}$ around 400\,kJ/m$^3$ and a saturation magnetization $J_\mathrm{s}$ of about 0.84\,T at 40 K. For these systems $K_\mathrm{eff}$ and $J_\mathrm{s}$ don't vary significantly with the repetition number $N$. This has already been shown at room temperature in Ref. \cite{heigl2020magnetic}.
The ferrimagnetic layer exhibits as well strong PMA with $K_\mathrm{eff}$ = 1000\,kJ/m$^3$ and $J_\mathrm{s}$ = 0.65\,T. As the magnetization of the ferrimagnet is dominated by the Tb magnetic moments at 40 K, strong antiferromagnetic coupling to the ferromagnetic layer is present \cite{schubert_interfacial_2013}. Before taking minor loops, the heterostructure sample is saturated at room temperature and cooled down in -7 T.
Easy-axis minor loops taken at 40 K of the FM/FI heterostructures are displayed in Fig.~\ref{fig:exp}. The magnetic moments of all films are normalized to that of the [Co(0.2\,nm)/Ni(0.4\,nm)/Pt(0.6\,nm)]$_{15}$ layer. Please note that the value of the remanent magnetization at zero field was arbitrary set to zero for better visibility. The magnetic moments of the different [Co/Ni/Pt] multilayers are plotted as a function of the applied negative field, with the magnetization of the ferrimagnetic TbCoFe layer being saturated and aligned with the negative field direction. Lowering the field from negative saturation will eventually reverse the magnetization of the ferromagnetic layer even before zero field which is driven by the strong antiferromagnetic coupling to the ferrimagnet \cite{schubert_interfacial_2013}. Afterwards, the field is increased again in negative field direction. For this applied field cycle the magnetization of the minor loop is recorded.  As shown in figure~\ref{fig:exp}, the field required for reversal becomes larger with decreasing number $N$. Note that this is expected for a constant ferro-/ferrimagnet interfacial exchange-coupling. However, below  a critical [Co/Ni/Pt] repetition number of $N<7$ the reversal becomes hysteresis free. The conditions necessary for this will be derived in the next section.

\section{Theoretical Modeling}

\subsection{Micromagnetic Simulations}
\label{sec:magnum}
In order to understand the underlying physics of the magnetization reversal process of these exchange-coupled bilayers, a finite-element software package magnum.fe~\cite{abert_magnum.fe:_2013} was used to simulate the field dependence of the total z-component of the magnetic moment by means of a spin-chain model. This model consists of a 3D nanorod with a square basal plane of side length $a=1$\,nm, but with a lateral discretization length much larger than 1\,nm. This produces a mesh with nodes only along the edges in lateral direction. Along the easy-axis direction (z direction) a fine mesh with a discretization length of approximately 0.75\,nm is used. The material parameters used for the micromagnetic simulations are partially based on the experimental data of the individual layers and are summarized in Table~\ref{tab:material}.
\begin{table}[h!]
  \centering
  \vspace{0.5cm}
  \begin{tabular}{c c c}
    \toprule
    \toprule
      & [Co/Ni/Pt]$_N$ & Tb$_{28}$Co$_{14}$Fe$_{58}$ \\
    \midrule
    $K_{\mathrm{eff}}$\,[kJ/m$^3$] & 151 & 1000\\
    $J_{\mathrm{S}}$\,[T] & 0.84 & 0.65 \\
    $A_{\mathrm{ex}}$\,[pJ/m] & 10.0 & 10.0 \\
    $J_{\mathrm{iex}}$\,[mJ/m$^2$] & \multicolumn{2}{c}{$-35.0$} \\
    $\lambda$ & 1.0 & 1.0 \\
    $\angle(\boldsymbol{K}_{\mathrm{eff}},\boldsymbol{e}_z)$\,[$^\circ$] & 1.0 & 1.0\\
    \midrule
    $a$\,[nm] & 1.0 & 1.0 \\
    $t$\,[nm] & 4.0 - 18.0 & 20.0\\
    \bottomrule
    \bottomrule
  \end{tabular}
  \caption{\small Material parameters used for the micromagnetic simulations of the investigated FM/FI heterostructure. $K_{\mathrm{eff}}$ is the effective anisotropy constant, $J_{\mathrm{S}}$ is the saturation magnetization, $A_{\mathrm{ex}}$ is the exchange coupling the bulk, $J_{\mathrm{iex}}$ is the interface exchange coupling between the antiferromagnetically coupled layers, $\lambda$ is the damping constant, $a$ is the side length of the square basal plane and $t$ the length of the 3D nanorod along the z direction (easy axis direction) used for the spin-chain model. The anisotropy axis is tilted by 1\,$^\circ$ against the z direction in both layers to avoid metastable states.}
  \label{tab:material}
\end{table}
\begin{figure}
    \centering
    \includegraphics[width=0.9\linewidth]{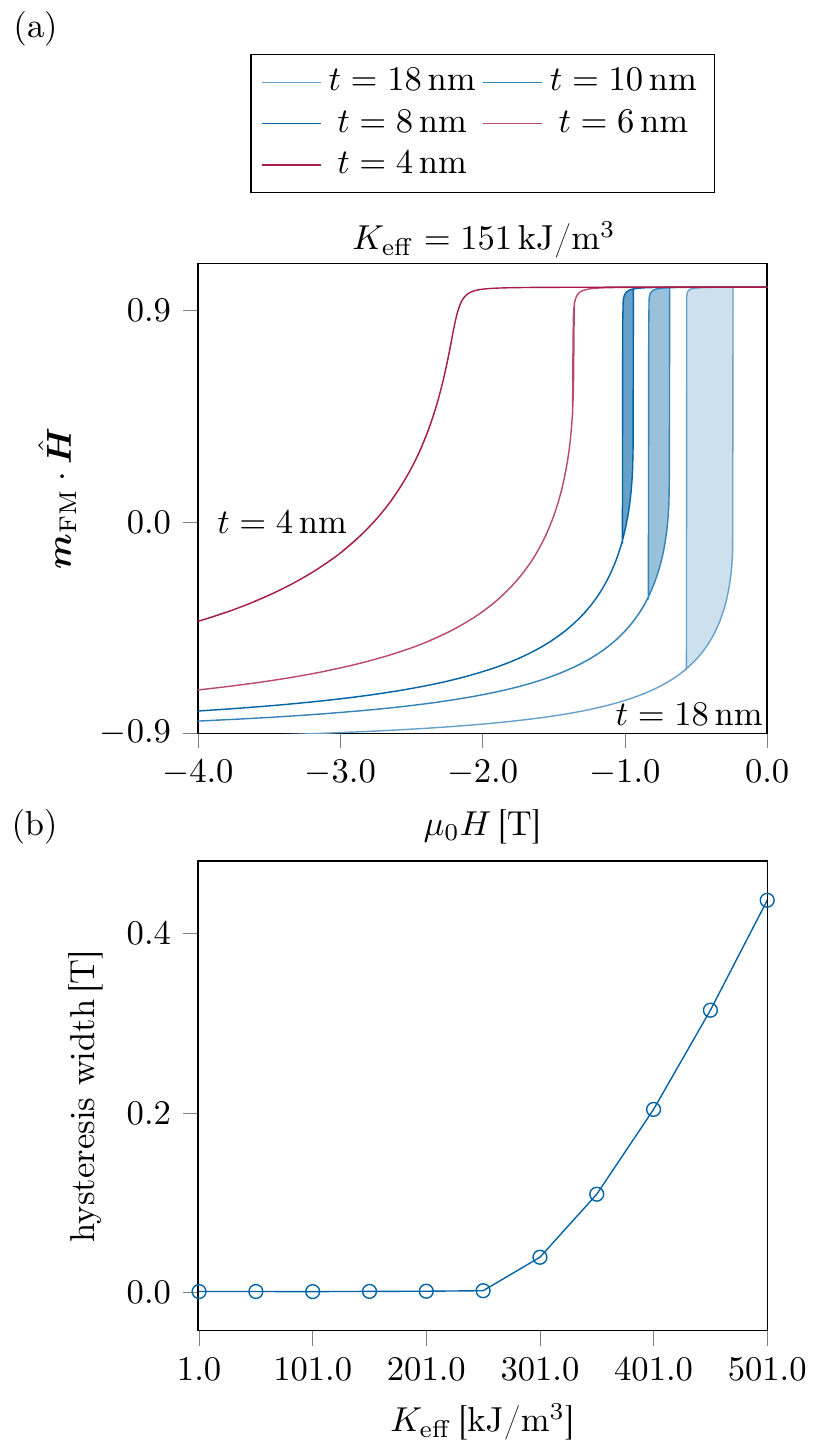}
    \caption{\small (color online) (a) Calculated easy-axis minor loops of the FM layers using $K_{\mathrm{eff}}=151$\,kJ/m$^3$ coupled to a 20\,nm thick TbCoFe layer for various thicknesses $t$ of the FM layer (used material parameters are given in Table~\ref{tab:material}). (b) Dependence of the corresponding hysteresis width on $K_{\mathrm{eff}}$ calculated for the heterostructure with $t$ = 4\,nm.}
    \label{fig:res1}
\end{figure}
In the micromagnetic simulations a single ferromagnetic layer with the properties of the [Co/Ni/Pt] multilayer is computed. Note that because of the selected vertical discretization length of approximately 0.75\,nm, the thicknesses of the modeled ferromagnetic layers are not multiples of the 1.2 nm period thickness of the [Co/Ni/Pt] stack.
The modeling is started with the zero-field state with the magnetization of the ferromagnetic layer pointing in +z direction and the TbCoFe net magnetization pointing in the -z direction. Then, the magnetic field magnitude is increased stepwise in $-2.5$\,mT increments up to -5\,T and back to 0\,T. After each field-step the micromagnetic state of the system is relaxed for 1\,ns. Note that the variation of the applied field in the modeling work is performed much faster than that used during the acquisition of the magnetometry data. However, because a high damping constant ($\lambda=1.0$) is used in the modeling work, a stationary state is obtained within 1\,ns, such that the modeled loops are representative for the experimental loops. Figure~\ref{fig:res1}(a) displays the minor hysteresis loops normalized by the saturation magnetization obtained from the spin-chain model for ferromagnetic layers with the magnetic anisotropy kept constant at 151\,kJ/m$^3$ and a film thickness varied between 4 and 18\,nm. The comparison of the modeled [Fig.~\ref{fig:res1}(a)] with the experimental results (Fig.~\ref{fig:exp}) reveals that the switching field (where most of the magnetic moment becomes aligned with the applied field) as well as the absence and presence of a hysteresis are well reproduced by our modeling work. As in the experiment, the reversal remains hysteresis free for FM layer thicknesses below a critical threshold thickness. Above this threshold value the minor loop becomes irreversible. 
However, the magnetic anisotropy of 151\,kJ/m$^3$ used in our calculations is reduced compared to the value determined experimentally for the [Co/Ni/Pt] reference samples.  
This deviation might be explained by the different growth conditions of the single reference layer grown on a 111-textured Pt seed layer promoting PMA \cite{doi:10.1063/1.4930830} compared to the heterostructures, where the ferromagnetic layer is deposited on amorphous TbFeCo.
The dependence of the hysteresis width on $K_{\mathrm{eff}}$ for 4\,nm thick FM layers is shown in  Fig.~\ref{fig:res1}(b). 
A minor loop hysteresis occurs only for anisotropies beyond a threshold anisotropy. Above this threshold anisotropy, the hysteresis width increases with $K_{\mathrm{eff}}$. This is reminiscent of the hysteresis width predicted by the Stoner-Wohlfarth theory. 
The micromagnetic simulations displayed in Fig.~\ref{fig:schematic}(a) reveal that a domain wall is formed across the interface to the TbCoFe layer during reversal, as will be discussed in more detail in the next section. 

\subsection{Analytical Model}
\label{sec:analytical}
In order to reproduce the micromagnetic spin-chain results and to find an analytical condition for the onset of the observed hysteresis-free minor loops, the following energy density is considered:
\begin{eqnarray}
\label{eq:energy_full}
 E=&-&\mu_0 H M_{\mathrm{FM}}t_{\mathrm{FM}} \cos(\alpha)\nonumber\\
 &+& K_{\mathrm{FM}}t_{\mathrm{FM}}\sin(\alpha)^2+ \frac{A_{\mathrm{FM}}}{t_{\mathrm{FM}}}(\beta-\alpha)^2\nonumber\\
 &+&|J_{\mathrm{iex}}|\cos\left(\gamma-\beta\right)\nonumber\\
 &+& 2\sqrt{A_{\mathrm{FI}}K_{\mathrm{FI}}} \left[1-\cos(\gamma-\pi)\right],
 \end{eqnarray}
\begin{figure}
      \centering
      \includegraphics[width=0.9\linewidth]{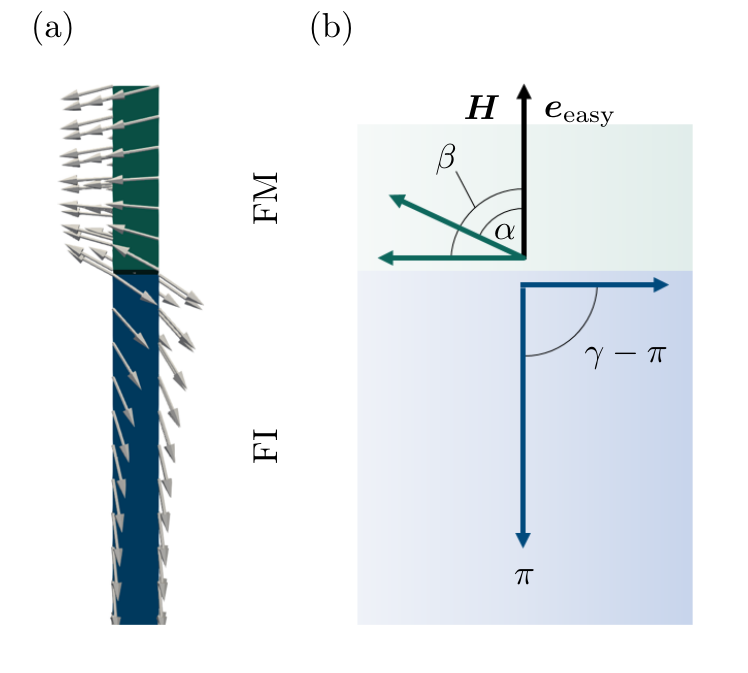}
      \caption{\small (color online) (a) Magnetization configuration of the micromagnetic spin-chain model at an external field of -2.35\,T ($m_{z,\mathrm{FM}}=0$) during the minor loop reversal illustrated in Fig.~\ref{fig:res1} for $K_{\mathrm{eff}}=201$\,kJ/m$^3$. Note that the bottom layer is cut off for better visibility. (b) Schematic illustration of the analytic model applied in Eq.~\ref{eq:energy_full}. $\mu_0\boldsymbol{H}$ is the external applied field, $\boldsymbol{e}_{\mathrm{easy}}$ is the uniaxial magnetic anisotropy axis of both layers, $\alpha$ is the angle of the average FM magnetization, and $\beta$ and $\gamma$ are the magnetization angles of the FM and the FI layers at the interface, respectively. All angles refer to $\boldsymbol{e}_{\mathrm{easy}}$ as origin. The ferromagnetic layer is shown in green and the ferrimagnetic layer is shown in blue.}
      \label{fig:schematic}  
\end{figure}
where the subscript FM denotes properties of the ferromagnet and FI those of the ferrimagnet; $M$ denotes the saturation magnetization, $K$ the effective magnetic anisotropy constant, $t$ the layer thickness, $A$ the bulk exchange and $|J_{\mathrm{iex}}|$ the interface exchange constant. As indicated in Fig.~\ref{fig:schematic}(b), $\alpha$ is the average magnetization angle of the FM and $\beta$ and $\gamma$ describe the maximum angles of the partial domain walls in the FM and in the FI layers, respectively. Further, we assume an external field $\mu_0\boldsymbol{H}$ applied along the easy-axis direction (z direction) of the bilayer structure. 

The presented analytical model treats the FM and the FI layer with different energy terms. The total energy of the FM consists of the Zeeman energy (first term) and the anisotropy energy (second term), which are based on its average magnetization direction. The third term of Eq.~\ref{eq:energy_full} describes the exchange interactions between neighboring spins within the FM, which is an important contribution to the domain wall energy. This energy term was derived in Ref.~\cite{goto_magnetization_1965} for the case of a soft magnetic layer with finite thickness  exchange-coupled to a ferromagnetic film with a higher coercivity. 


The bottom layer of our structure is a ferrimagnet with a large thickness. Large means that we assume the FI to be much thicker than the domain wall width, or in other words, the FI must be sufficiently thick that the spins at the bottom end point along the easy axis [see. Fig.~\ref{fig:schematic}(a)]. If we further assume that the anisotropy field of the FI is large compared to the applied external field, the zero-field approximation of Ref.~\cite{zijlstra_coping_1979} can be used for the FI. This energy term (last term in Eq.~\ref{eq:energy_full}) describes the energy associated to the partial domain wall occurring inside the ferrimagnet. The angle $\gamma-\pi$ describes the deviation of the magnetic moments away from the down direction that is largest at the FI/FM interface and gradually decays away from this interface inside the ferrimagnet. The expression is valid if the magnetic moments at the ferrimagnet bottom are aligned parallel to the down direction. Conveniently, the contributions of the Zeeman energy and the anisotropy energy are already included in the last term of Eq.~\ref{eq:energy_full}.
Note, that such an energy term is also used in the Mauri model~\cite{mauri_simple_1987} to describe the rotation of the antiferromagnetic spins arising from the exchange torque based on the rotation of the FM and its bidirectional coupling to the antiferromagnet. The fourth term in Eq.~\ref{eq:energy_full} is an interface exchange of the Heisenberg type, which describes the antiferromagnetic coupling of the neighboring spins at the FI/FM interface.

To derive a condition for hysteresis-free switching, in principle a full stability analysis of Eq.~\ref{eq:energy_full} is necessary in order to obtain the equilibrium angles of the moments in the FM $\alpha$ and the angles describing the partial domain walls $\beta$ and $\gamma$ in the FM and FI layers, respectively. Without further assumptions a stability analysis can only be done numerically. In the appendix we derive how to reduce Eq.~\ref{eq:energy_full} to an approximate function depending on the variable $\alpha$ solely, by assuming strong interface exchange coupling between the FM and the FI layer:
\begin{eqnarray}
 \label{eq:energy_reduced_alpha}
  E(\alpha)=&-&\mu_0 H M_{\mathrm{FM}}t_{\mathrm{FM}} \cos(\alpha)\nonumber\\
  &+& K_{\mathrm{FM}}t_{\mathrm{FM}}\sin(\alpha)^2\nonumber\\
  &+& \frac{\frac{\sigma_{\mathrm{FM}}}{2}-\sigma_{\mathrm{FM}}k+\frac{\sigma_{\mathrm{FI}}}{2}k^2\left( 1+\frac{\sigma_{\mathrm{FM}}}{|J_{\mathrm{iex}}|} \right)}{\left( 1+\frac{\sigma_{\mathrm{FM}}}{|J_{\mathrm{iex}}|} \right)}\alpha^2,
\end{eqnarray}
with:
\begin{equation}
 k=\frac{\frac{\sigma_{\mathrm{FM}}}{|J_{\mathrm{iex}}|}}{\left( 1+\frac{\sigma_{\mathrm{FM}}}{|J_{\mathrm{iex}}|} \right) \left( 1+\frac{\sigma_{\mathrm{FI}}}{|J_{\mathrm{iex}}|}-\frac{1}{1+\frac{\sigma_{\mathrm{FM}}}{|J_{\mathrm{iex}}|}}\right)}.
\end{equation}
To shorten the notation, the variables $\sigma_{\mathrm{FM}}=2A_{\mathrm{FM}}/t_{\mathrm{FM}}$ and $\sigma_{\mathrm{FI}}=2\sqrt{A_{\mathrm{FI}}K_{\mathrm{FI}}}$ were introduced.

\subsection{Stoner-Wohlfarth Interpretation}
\label{sec:SW}
\begin{figure}
    \centering
    \includegraphics[width=0.9\linewidth]{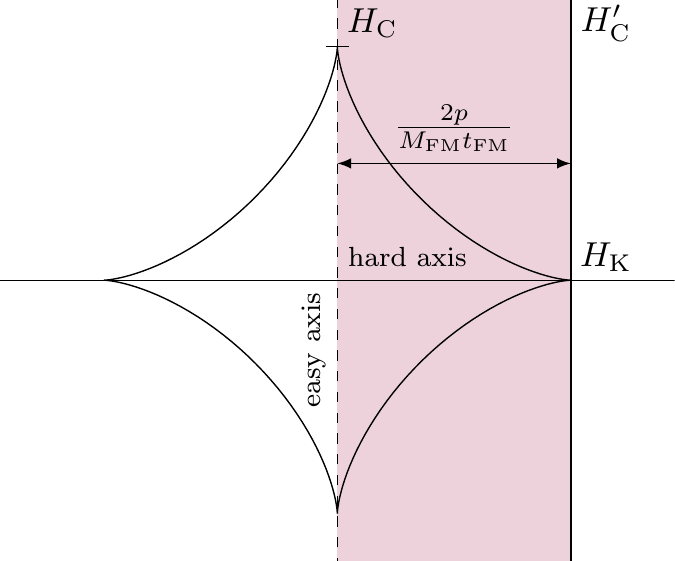}
\caption{\small (color online) Stoner-Wohlfarth astroid with an intrinsic hard-axis bias field $\frac{2p}{M_{\mathrm{FM}}t_{\mathrm{FM}}}=H_{\mathrm{K}}$.}
\label{fig:SW_astroid}
\end{figure}

To obtain a condition for hysteresis-free switching and gain insight into the physical mechanism leading to it, Eq.~\ref{eq:energy_reduced_alpha} is converted to a form allowing a direct comparison with the energy density terms of a Stoner-Wohlfarth (SW) particle. With this, all findings from the SW theory can be applied to the model presented here. In particular, the stability analysis and the associated energy minima based on the material properties and the external applied field should be mentioned here. As can be seen in Fig.~\ref{fig:res1}(a), the magnetization behavior of the FM layer in the range around $\alpha=\pi/2$ needs to be considered for evaluating whether hysteresis-free switching occurs. Transforming the coordinates with $\alpha=\alpha_0-\epsilon$ of Eq.~\ref{eq:energy_reduced_alpha}, with $\alpha_0=\pi/2$ and small $\epsilon$, Eq.~\ref{eq:energy_reduced_alpha} becomes:
\begin{eqnarray}
\label{eq:energy_reuced}
  E=&-&\mu_0 H M_{\mathrm{FM}}t_{\mathrm{FM}} \sin(\epsilon)\nonumber\\
 &+& K_{\mathrm{FM}}t_{\mathrm{FM}}\cos(\epsilon)^2+p\alpha_0^2-2p\alpha_0\epsilon+p\epsilon^2.
\end{eqnarray}
Constant terms like $p\alpha_0^2$ just shift the zero point energy $E\rightarrow E^\ast$. For small $\epsilon$ the linear term can be interpreted as $\epsilon\sim\sin(\epsilon)$ and the quadratic term can be interpreted as $\epsilon^2\sim2(1-\cos(\epsilon))$. With these approximations, Eq.~\ref{eq:energy_reuced} becomes:
\begin{eqnarray}
\label{eq:energy_SW}
   E^\ast = &\phantom{-}&K_{\mathrm{FM}}t_{\mathrm{FM}}\cos(\epsilon)^2\nonumber\\
   &-& M_{\mathrm{FM}}t_{\mathrm{FM}}
            \begin{pmatrix}
                \mu_0H+p\pi\\
                \frac{2p}{M_{\mathrm{FM}}t_{\mathrm{FM}}}
            \end{pmatrix}
            \cdot
            \begin{pmatrix}
                \sin(\epsilon) \\
                \cos(\epsilon)
            \end{pmatrix}.
\end{eqnarray}
If we compare this form with the energy density of a classical SW particle with an angle around $\pi/2$ (see appendix) Eq.~\ref{eq:energy_SW} can be identified as the energy density of such a SW particle with a small angle $\epsilon$ around $\alpha_0=\pi/2$, with an effective field with a component $\mu_0 H^\ast=\mu_0H+p\pi$ in the easy-axis direction and a component $2p/(M_{\mathrm{FM}}t_{\mathrm{FM}})$ in the hard-axis direction. As mentioned, all findings from the SW theory can now be applied to the presented model. In particular, we know from the SW astroid (see Fig.~\ref{fig:SW_astroid}) that there can only exist two energy minima (irreversible switching) for fields in the hard-axis direction that are lower than the anisotropy field:
\begin{equation}
 \frac{2p}{M_{\mathrm{FM}}t_{\mathrm{FM}}}<\frac{2K_{\mathrm{FM}}}{M_{\mathrm{FM}}}.
\end{equation}
From this requirement we directly obtain a condition for hysteresis-free switching:
\begin{equation}
\label{eq:cond}
 \frac{\frac{\sigma_{\mathrm{FM}}}{2}-\sigma_{\mathrm{FM}}k+\frac{\sigma_{\mathrm{FI}}}{2}k^2\left( 1+\frac{\sigma_{\mathrm{FM}}}{|J_{\mathrm{iex}}|} \right)}{\left( 1+\frac{\sigma_{\mathrm{FM}}}{|J_{\mathrm{iex}}|} \right)}>K_{\mathrm{FM}}t_{\mathrm{FM}}.
\end{equation}
In addition, the left-hand side of Eq.~\ref{eq:cond} can be interpreted as an intrinsic hard-axis bias field that arises from the partial domain walls in the two layers of the investigated FM/FI structure.

\section{Results}
\label{sec:res}
\begin{figure}
    \centering
    \includegraphics[width=0.9\linewidth]{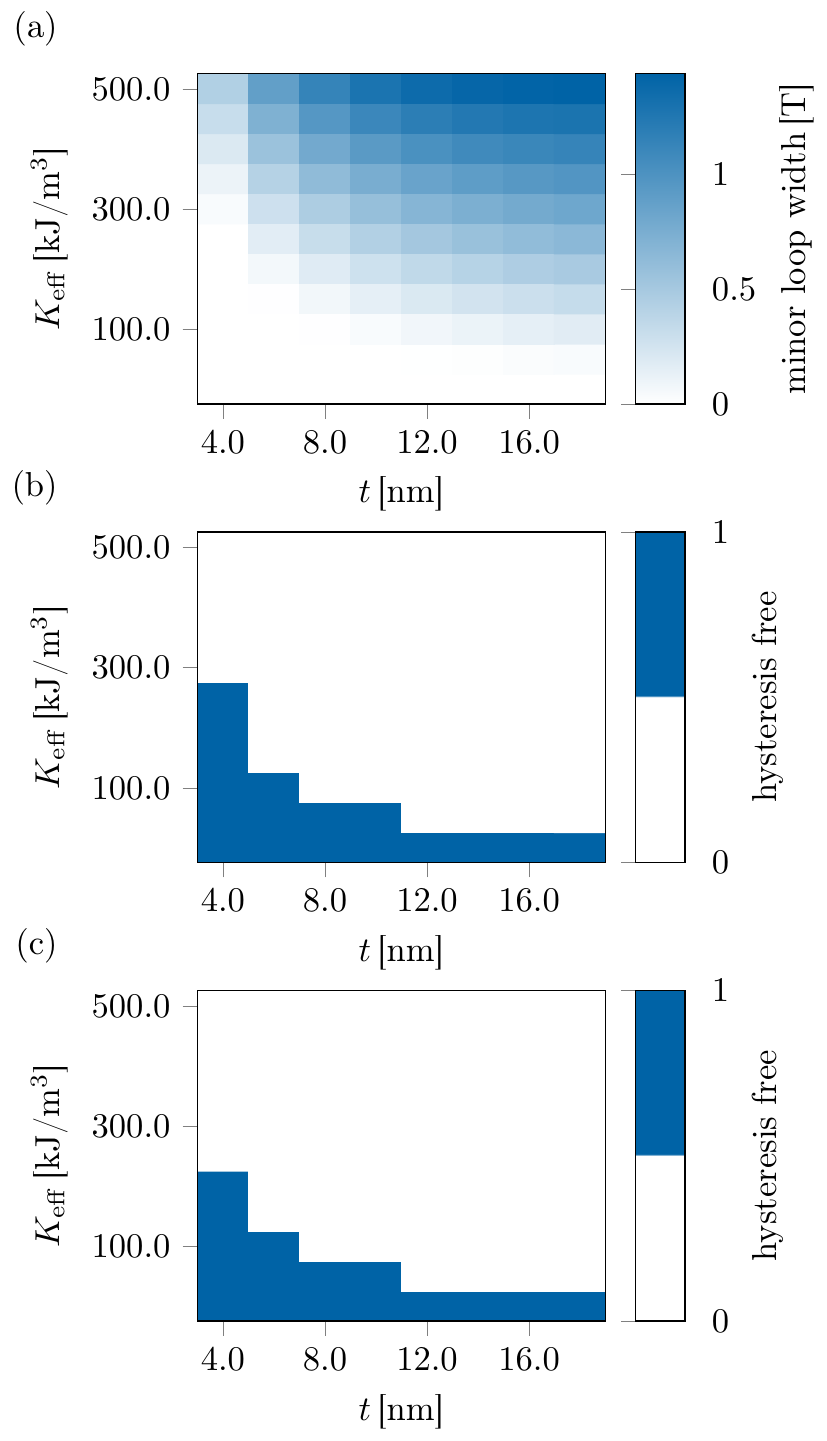}
    \caption{\small (color online) (a) Hysteresis width of easy-axis minor loops of a FM layer antiferromagnetically coupled to a 20\,nm thick FI Tb$_{28}$Co$_{14}$Fe$_{58}$ layer for various effective uniaxial magnetic anisotropy constants and thicknesses of the FM layer (all other material parameters are shown in Table~\ref{tab:material}). (b) Binary phase diagram of (a) with a threshold hysteresis width of 2.5\,mT. (c) Binary phase diagram of the parameter space of (b) based on the condition of Eq.~\ref{eq:cond}.}
    \label{fig:res3}
\end{figure}
\begin{figure}
    \centering
    \includegraphics[width=0.9\linewidth]{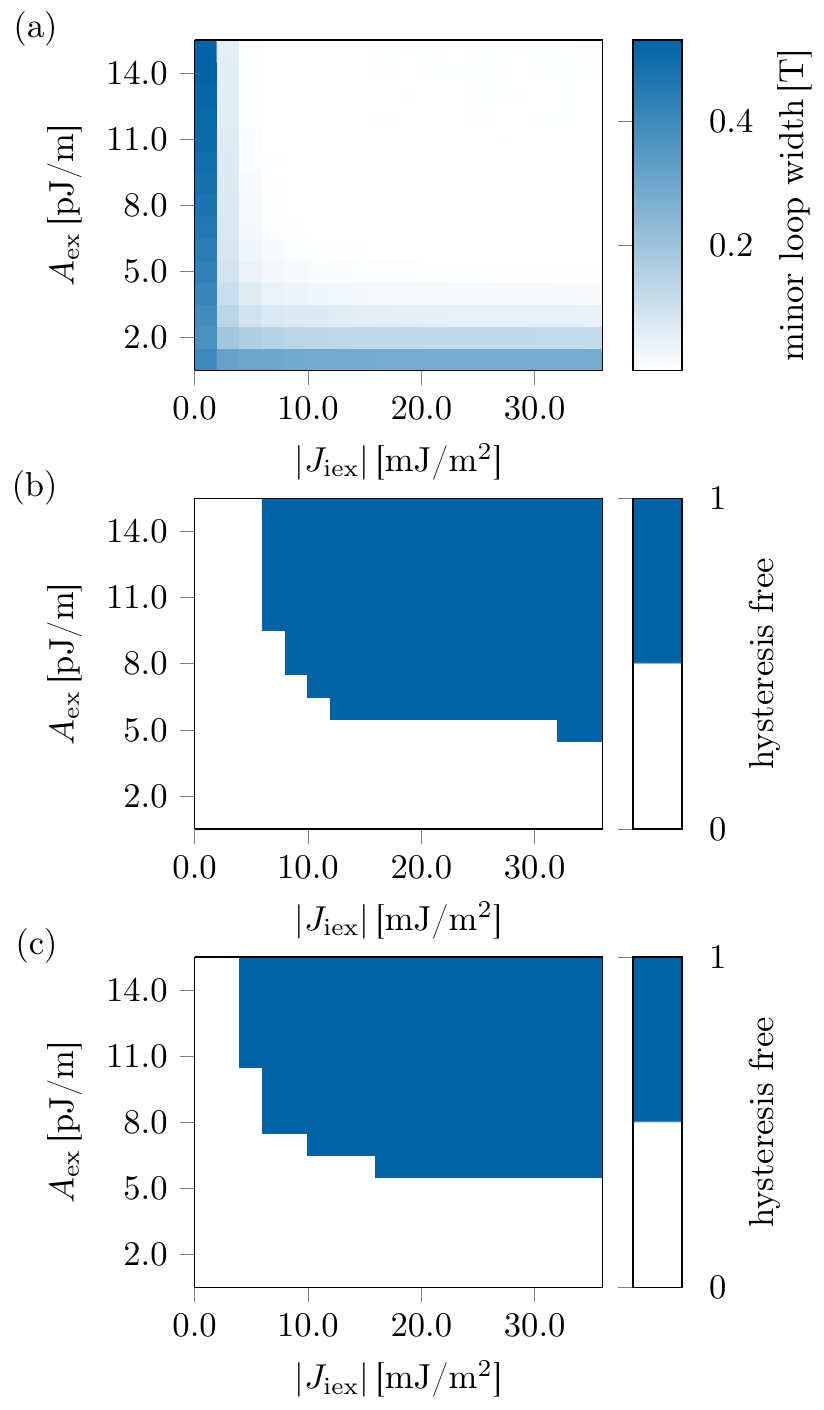}
    \caption{\small (color online) (a) Hysteresis width of easy-axis minor loops of a FM layer antiferromagnetically coupled to a 20\,nm thick FI Tb$_{28}$Co$_{14}$Fe$_{58}$ layer for various interface exchange constants and for various bulk exchange constants of the FM layer (all other material parameters are shown in Table~\ref{tab:material}). (b) Binary phase diagram of (a) with a threshold hysteresis width of 2.5\,mT. (c) Binary phase diagram of the parameter space of (b) based on the condition of Eq.~\ref{eq:cond}.}
    \label{fig:res4}
\end{figure}
\begin{figure}
    \centering
    \includegraphics[width=0.9\linewidth]{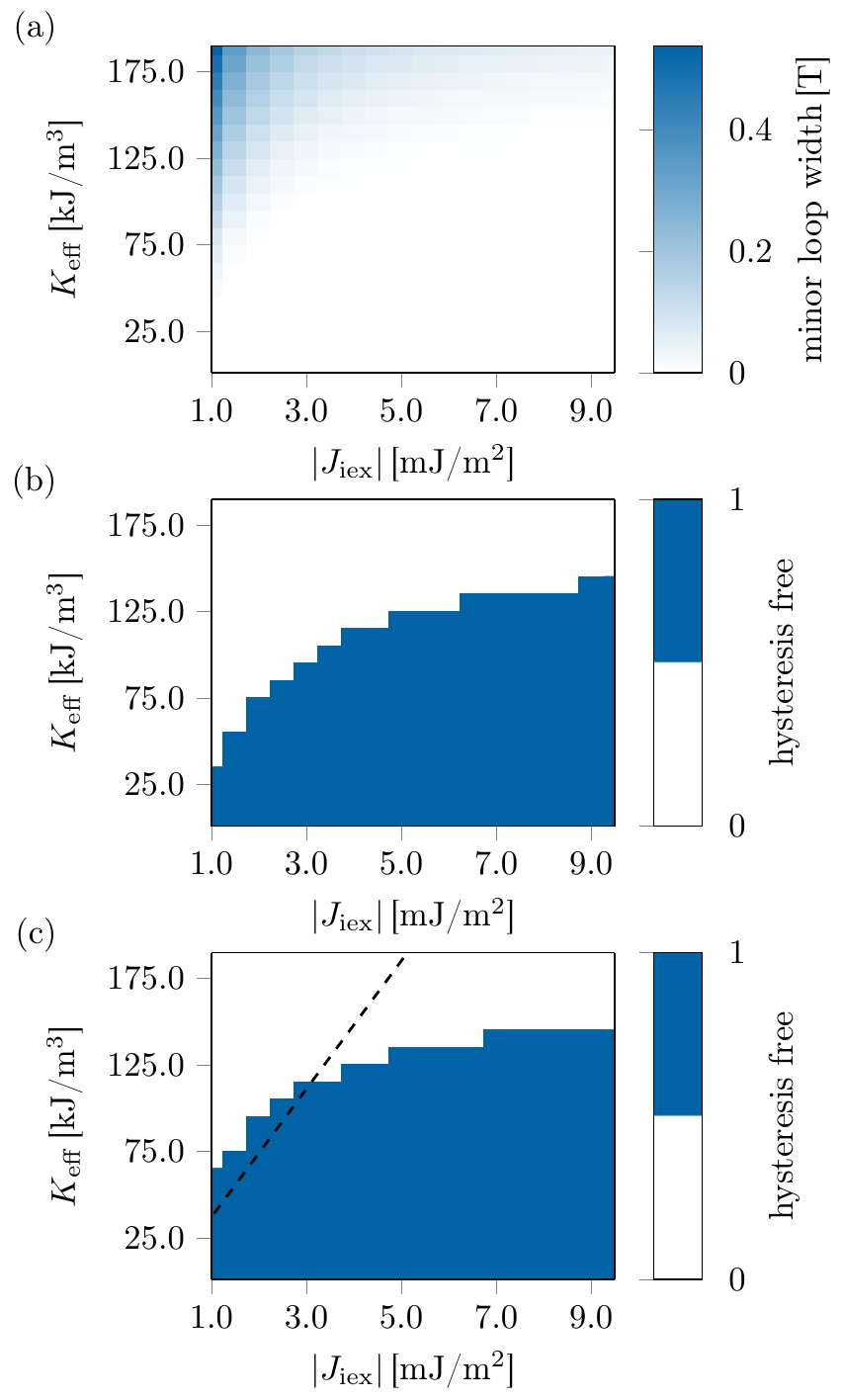}
    \caption{\small (color online) (a) Hysteresis width of easy-axis minor loops of a 5.5\,nm thick FM layer antiferromagnetically coupled to a 20\,nm thick FI Tb$_{28}$Co$_{14}$Fe$_{58}$ layer for various interface exchange constants and effective uniaxial magnetic anisotropy constants of the FM layer (all other material parameters are shown in Table~\ref{tab:material}). (b) Binary phase diagram of (a) with a threshold hysteresis width of 2.5\,mT. (c) Binary phase diagram of the parameter space of (b) based on the condition of Eq.~\ref{eq:cond}. The dashed line indicates a constant ratio $|J_{\mathrm{iex}}|/K_{\mathrm{eff}}=0.027$\,$\mu$m (as was assumed in Ref.~\cite{zhao_magnetization_2019}).}
    \label{fig:res5}
\end{figure}
To test the validity of Eq.~\ref{eq:cond}, we perform spin-chain simulations as described in Section~\ref{sec:magnum}. First, the effective  magnetic anisotropy constant of the FM is varied in the range of $1$\,kJ/m$^3$ - $501$\,kJ/m$^3$ with $\Delta K_{\mathrm{eff}}=50$\,kJ/m$^3$ and the layer thickness is varied in the range of $4$\,nm - $18$\,nm with $\Delta t=2$\,nm. All other parameters are given in Table~\ref{tab:material}. Figure~\ref{fig:res3}(a) shows a phase diagram of the resulting width of the minor hysteresis loop. If the points with a hysteresis width of more and less than 2.5\,mT (resolution of the loop) are separated, a binary diagram can be obtained, as illustrated in Fig.~\ref{fig:res3}(b). Such a diagram separates $K_{\mathrm{eff}}$,$t_\mathrm{{FM}}$-regions with hysteretic switching from those that switch without hysteresis. We can now compare these binary diagrams with the condition for hysteresis-free switching obtained from the analytical model (Eq.~\ref{eq:cond}), shown in Figure~\ref{fig:res3}c. The good agreement of the numerical modeling results with the analytical model confirms the validity of the latter for the chosen parameter range.

In a second step, $|J_{\mathrm{iex}}|$ is varied from weak to full exchange in the range of 1\,mJ/m$^2$ - 35\,mJ/m$^2$ with $\Delta |J_{\mathrm{iex}}|=2$\,mJ/m$^2$ and the ferromagnetic bulk exchange $A_{\mathrm{ex}}$ is varied in the range of 1\,pJ/m$^3$ - 15\,pJ/m$^3$ with $\Delta A_{\mathrm{ex}}=1$\,pJ/m$^3$. Note, that the properties of the FI bottom layer remain unchanged. The resulting phase diagram is displayed in Fig.~\ref{fig:res4}(a) and the binary separation is shown in Fig.~\ref{fig:res4}(b) together with the corresponding diagram based on the analytical prediction in Fig.~\ref{fig:res4}(c). Again an excellent agreement between the numerically obtained results and the analytical model is apparent. The latter correctly predicts hysteresis-free switching for high $|J_{\mathrm{iex}}|$ and high $A_{\mathrm{ex}}$. In addition, qualitatively the border of hysteresis-free switching shifts to higher values of $A_{\mathrm{ex}}$ for decreasing $|J_{\mathrm{iex}}|$ coming from strong interface exchange.  

Last, the dependence of the magnetization reversal process on the FM anisotropy and interface exchange is investigated. Here, $K_{\mathrm{eff}}$ is varied from $1$\,kJ/m$^3$ to $191$\,kJ/m$^3$ with $\Delta K_{\mathrm{eff}}=10$\,kJ/m$^3$, and $|J_{\mathrm{iex}}|$ is varied from 1\,pJ/m$^3$ - 9.5\,pJ/m$^3$ with $\Delta |J_{\mathrm{iex}}|=0.5$\,mJ/m$^2$. As shown in Fig.~\ref{fig:res5}, also with these parameters the results of the micromagnetic simulations and the analytical condition agree extremely well. Interestingly, the data plotted in Figs.~\ref{fig:res5}(a) and (b) explain why a spatial variation of $K_{\mathrm{eff}}$ in Ref.~\cite{zhao_magnetization_2019} also required a respective variation of $|J_{\mathrm{iex}}|$ in order to observe hysteresis-free switching in large areas of the investigated [Co(0.4\,nm)/Pt(0.7\,nm)]$_5$/Tb$_{26.5}$Co$_{73.5}$ thin FM/FI bilayer system. In the latter work the ratio $|J_{\mathrm{iex}}|/K_{\mathrm{eff}}$ was kept constant at 0.027\,$\mu$m which corresponds to the dashed line in Fig.~7(c). Although, the phase boundary is not a straight line, the dependence between $J_{\mathrm{iex}}$ and $K_{\mathrm{eff}}$ chosen in Ref.~\cite{zhao_magnetization_2019} approximates the phase boundary between a hysteretic and hysteresis-free minor magnetization reversal reasonably well. 


\section{Discussion}
The good agreement of the analytical prediction with the micromagnetic spin-chain model for large interface exchange in Fig.~\ref{fig:res4} is noteworthy, but much more surprising is the agreement for small $|J_{\mathrm{iex}}|$, since the condition for vanishing hysteresis, i.e. Eq.~\ref{eq:cond} in Sec.~\ref{sec:analytical} was derived under the condition of a strong interface exchange. As a consequence, some limiting cases of Eq.~\ref{eq:cond} for weak interface exchange coupling are investigated in the following.

First, we consider the case of $\sigma_{\mathrm{FI}}/|J_{\mathrm{iex}}|\gg1$ and $\sigma_{\mathrm{FM}}/|J_{\mathrm{iex}}|\gg1$. Since $\sigma_{\mathrm{FM}}$ and $\sigma_{\mathrm{FI}}$ are a measure of the energy cost of a domain wall in the FM and FI layers, the formation of partial domain walls in both layers is very expensive. As a consequence, the FM always switches irreversibly from an antiparallel state at zero field to a parallel state at high fields, without forming an interfacial domain wall. In this limit $k\rightarrow |J_{\mathrm{iex}}|/\sigma_{\mathrm{FI}}$ and Eq.~\ref{eq:cond} reduces to:
\begin{equation}
 \frac{|J_{\mathrm{iex}}|}{2}>K_{\mathrm{FM}}t_{\mathrm{FM}}.
\end{equation}
For small $|J_{\mathrm{iex}}|$, the hysteresis-free switching condition is solely determined by the interface exchange coupling strength. For vanishing $|J_{\mathrm{iex}}|$, hysteresis-free switching is only possible for vanishing anisotropy.

Second, we still consider $\sigma_{\mathrm{FI}}/|J_{\mathrm{iex}}|\gg1$, but this time $\sigma_{\mathrm{FM}}\sim|J_{\mathrm{iex}}|$. This means it is energetically much less favorable to form a domain wall in the FI than in the FM. In contrast to the previous limit, $|J_{\mathrm{iex}}|$ is large enough that the interface exchange cannot be overcome by the FM reversal field. Hence, a partial domain wall develops in the FM, but not in the FI. In this limit $k\rightarrow |J_{\mathrm{iex}}|/(2\sigma_{\mathrm{FI}})$ and Eq.~\ref{eq:cond} becomes:
\begin{equation}
 \frac{|J_{\mathrm{iex}}|}{4}>K_{\mathrm{FM}}t_{\mathrm{FM}}.
\end{equation}

These limiting cases show that Eq.~\ref{eq:cond}, the condition for hysteresis-free switching, although derived under the condition of strong interface exchange, still correctly reproduces the behavior for weak $|J_{\mathrm{iex}}|$. In addition, we get a further confirmation of the interpretation of Sec.~\ref{sec:SW}. Hysteresis-free switching only occurs if partial domain walls in one or both of the layers can develop such that the Zeeman energy provided by the external field becomes stored as exchange energy in the partial wall(s). These must contain sufficient energy that the arising intrinsic hard-axis field is larger than the anisotropy field of the FM, which is equivalent to the reduction of the energy barrier of the FM reversal along the easy-axis direction to zero.

A further interesting limiting case for strong interface exchange coupling is $\sigma_{\mathrm{FM}}/|J_{\mathrm{iex}}|\ll1$ and $\sigma_{\mathrm{FI}}/|J_{\mathrm{iex}}|\ll1$ with $\sigma_{\mathrm{FM}}/\sigma_{\mathrm{FI}}\ll1$. This means we have strong interface exchange coupling, and a partial domain wall in the FI is energetically much less favourable than in the FM. For this case, we obtain $k\rightarrow \frac{\sigma_{\mathrm{FM}}}{\sigma_{\mathrm{FI}}}$ and the hysteresis-free switching condition:
\begin{equation}
 \frac{\sigma_{\mathrm{FM}}}{2}=\frac{A_{\mathrm{FM}}}{t_{\mathrm{FM}}}>K_{\mathrm{FM}}t_{\mathrm{FM}},
\end{equation}
which is equivalent to:
\begin{equation}
 \sqrt{\frac{A_{\mathrm{FM}}}{K_{\mathrm{FM}}}}>t_{\mathrm{FM}}.
\end{equation}
We see that the thickness of the FM must be sufficiently small to suppress the formation of a full domain wall inside the ferromagnetic layer. Moments in the hard-axis direction at the interface then produce a large intrinsic hard-axis bias field during reversal, which is a very reasonable result. Note that this limiting case was investigated in Ref.~\cite{wang_transition_2011} with simulations and the same dependency was reported. 

It should be noted that the existence of such types of hysteresis-free minor loops are not restricted to antiferromagnetically coupled bilayers. Even for pure ferromagnetic exchange-spring structures consisting of a soft and a hard magnetic layer the analytical model of Sec.~\ref{sec:analytical} can be applied and the results hold. Such structures have already been extensively investigated both theoretically and experimentally in the context of the development of permanent magnets~\cite{meiklejohn_new_1956,fullerton_exchange-spring_1998,fullerton_hard/soft_1999,kang_recoil_2005,william_mccallum_requirements_2006,lyubina_two-particle_2010} or grains for magnetic recording media~\cite{suess_exchange_2005-1,suess_exchange_2005,suess_multilayer_2006}. But in the SW-like models the hard-axis bias field of a 180 degree domain wall, that typically forms in the soft magnetic layer and gets pinned at the interface to the hard magnet, was not taken into account. Based on this domain wall, an intrinsic hard-axis bias field occurs and hysteresis-free switching is possible. To support this hypothesis, we refer to the work of Fullerton~et~al.~\cite{fullerton_exchange-spring_1998}, in which the authors have investigated exchange-coupled SmCo/Fe and SmCo/Co bilayers with thicknesses of Fe and Co of 10\,nm and 20\,nm, respectively. Among others, the authors have studied easy-axis minor loops of these structures and found that the SmCo/Fe bilayer with 20\,nm Fe shows a hysteresis-free minor loop, but not that of the SmCo/Co bilayer with the same Co thickness. From the point of view of this work, this behavior can be easily explained by means of the condition of Eq.~\ref{eq:cond}. If the material parameters of the original work for Fe and Co are used only in the case of the 20\,nm Fe layer, the product of uniaxial magnetic anisotropy and layer thickness is sufficiently small that the bias field of the domain wall at the interface to SmCo can overcome the anisotropy field. This is true if, instead of $K_{\mathrm{FM}}=0.1$\,kJ/m$^3$, a still realistic Fe anisotropy constant of up to $K_{\mathrm{FM}}=59$\,kJ/m$^3$ is used. In contrast, the anisotropy constant of the Co layer with the same thickness is already too large to fulfill Eq.~\ref{eq:cond}, yielding a finite width of the minor loop.

\section{Conclusion}
In this work, we have shown experimentally and with micromagnetic simulations that hysteresis-free minor loops of a ferromagnetic top layer with finite magnetic anisotropy ([Co(0.2\,nm)/Ni(0.4\,nm)/Pt(0.6\,nm)]$_N$) exchange-coupled to a ferrimagnetic layer (Tb$_{28}$Co$_{14}$Fe$_{58}$) can exist. Furthermore, an analytical model was derived and based on it a condition for hysteresis-free (fully reversible) minor loop reversal of the ferromagnetic layer. The expression of the systems energy density approximated from the analytical model is reminiscent to the description of a magnetization reversal based on the Stoner-Wohlfarth model. With this comparison a fundamental understanding of the physics relevant for a hysteresis-free magnetization reversal of the FM could be obtained: 
Reversible loops always occur if a partial domain wall is formed inside the ferromagnetic and ferrimagnetic layer during the reversal of the FM. This domain wall generates an intrinsic hard-axis bias field that can overcome the anisotropy field of the FM, which makes the minor loop completely reversible. 
Although our analytical model was derived for the strong coupling regime, we have additionally shown by means of limiting cases that the presented condition is also applicable for weak coupling. 
Finally, we note that our analytical model is not restricted to antiferromagnetically coupled bilayers but is also valid for structures consisting of ferromagnetically coupled soft and a hard magnetic layers. We thus conclude that the analytical model derived here will be beneficial for the analysis of the magnetization reversal of exchange-coupled structures used in various applications.

\section{Acknowledgement}
Financial support for this joint D.A.CH project was provided by the German Research Foundation under Grant AL 618/24-1, the Swiss National Science Foundation under Grant 200021-147084, and the Austrian Science Fund under Grant I2214-N20, and is gratefully acknowledged.

\appendix
\section{Stability Analysis}
\label{sec:stability}
In equilibrium the partial derivatives of the energy of Eq.~\ref{eq:energy_full} with respect to $\beta$ and $\gamma$ must vanish:
\begin{eqnarray}
\label{eq:energy_3spin_P_beta}
 \frac{\partial E}{\partial \beta}&=&\sigma_{\mathrm{FM}}(\beta-\alpha)+|J_{\mathrm{iex}}|\sin(\gamma-\beta)=0\\
 \label{eq:energy_3spin_P_gamma}
 \frac{\partial E}{\partial \gamma}&=&-|J_{\mathrm{iex}}|\sin(\gamma-\beta)+\sigma_{\mathrm{FI}}\sin(\gamma-\pi)=0.
\end{eqnarray}
Here, the new variables $\sigma_{\mathrm{FM}}=2A_{\mathrm{FM}}/t_{\mathrm{FM}}$ and $\sigma_{\mathrm{FI}}=2\sqrt{A_{\mathrm{FI}}K_{\mathrm{FI}}}$ are used to shorten the notation. Under the assumption of strong interface exchange coupling, $\gamma-\beta$ can be approximated by $\pi$. This permits a Taylor expansion to the linear term of the sine in Eq.~\ref{eq:energy_3spin_P_beta} around $\pi$:
\begin{equation}
 \label{eq:3spin_beta_1}
 \sigma_{\mathrm{FM}}(\beta-\alpha)=|J_{\mathrm{iex}}|(\gamma-\beta-\pi).
\end{equation}
With this, $\beta$ becomes:
\begin{equation}
 \label{eq:3spin_beta_2}
 \beta=\frac{\frac{\sigma_{\mathrm{FM}}}{|J_{\mathrm{iex}}|}\alpha+(\gamma-\pi)}{1+\frac{\sigma_{\mathrm{FM}}}{|J_{\mathrm{iex}}|}}.
\end{equation}
The same expansion can be done for Eq.~\ref{eq:energy_3spin_P_gamma}. Additionally, if $\gamma-\beta\rightarrow\pi$ also $\gamma-\pi$ must become sufficiently small that Eq.~\ref{eq:energy_3spin_P_gamma} has a solution. Hence, also the second sine in Eq.~\ref{eq:energy_3spin_P_gamma} can be linearized, and equation Eq.~\ref{eq:energy_3spin_P_gamma} can be approximated by:
\begin{equation}
 \label{eq:3spin_gamma_1}
 |J_{\mathrm{iex}}|(\gamma-\beta-\pi)+\sigma_{\mathrm{FI}}(\gamma-\pi)=0.
\end{equation}
With this we obtain a linear equation in $\beta$ and $\gamma$:
\begin{equation}
 \label{eq:3spin_gamma_2}
 \left(1+\frac{\sigma_{\mathrm{FI}}}{|J_{\mathrm{iex}}|}\right)(\gamma-\pi)-\beta=0
\end{equation}
After inserting $\beta$ in Eq.~\ref{eq:3spin_gamma_1} a linear relation between $\gamma$ and $\alpha$ is obtained:
\begin{equation}
 \label{eq:3spin_gamma}
 \gamma-\pi=\frac{\frac{\sigma_{\mathrm{FM}}}{|J_{\mathrm{iex}}|}}{\left( 1+\frac{\sigma_{\mathrm{FM}}}{|J_{\mathrm{iex}}|} \right) \left( 1+\frac{\sigma_{\mathrm{FI}}}{|J_{\mathrm{iex}}|}-\frac{1}{\frac{\sigma_{\mathrm{FM}}}{|J_{\mathrm{iex}}|}}\right)}\alpha=k\alpha
\end{equation}
This equation can then be used in Eq.~\ref{eq:3spin_beta} to express also $\beta$ with just $\alpha$:
\begin{equation}
 \label{eq:3spin_beta}
 \beta=\frac{\frac{\sigma_{\mathrm{FM}}}{|J_{\mathrm{iex}}|}+k}{1+\frac{\sigma_{\mathrm{FM}}}{|J_{\mathrm{iex}}|}}\alpha.
\end{equation}
The last step is now to eliminate $\beta$ and $\gamma$ in Eq.~\ref{eq:energy_full}. If we again use the assumption of strong interface exchange coupling and make a Taylor expansion of $\cos(\gamma-\beta-\pi)$ and $\cos(\gamma-\pi)$ up to the quadratic terms, after lengthy but simple algebraic rearrangements, the total energy is reduced to:
\begin{eqnarray}
  E(\alpha)=&-&\mu_0 H M_{\mathrm{FM}}t_{\mathrm{FM}} \cos(\alpha)+c\nonumber\\
  &+& K_{\mathrm{FM}}t_{\mathrm{FM}}\sin(\alpha)^2\nonumber\\
  &+& \frac{\frac{\sigma_{\mathrm{FM}}}{2}-\sigma_{\mathrm{FM}}k+\frac{\sigma_{\mathrm{FI}}}{2}k^2\left( 1+\frac{\sigma_{\mathrm{FM}}}{|J_{\mathrm{iex}}|} \right)}{\left( 1+\frac{\sigma_{\mathrm{FM}}}{|J_{\mathrm{iex}}|} \right)}\alpha^2\nonumber\\
\end{eqnarray}
Here, $c$ includes all constant terms that appear during the algebraic rearrangements. This constant can be interpret as a shift of the zero point energy.

\section{SW particle around $\pi/2$}
\label{sec:classicalSW}
The classical Stoner-Wohlfarth energy density per unit area of a particle subject to an external field with a component in easy-axis direction $\mu_0 H_{\mathrm{ea}}$ and a component in hard-axis direction $\mu_0 H_{\mathrm{ha}}$ is:
\begin{equation}
\label{eq:energy_SW_classical}
   E = Kt\sin( \alpha)^2 - Mt
            \begin{pmatrix}
                \mu_0 H_{\mathrm{ea}}\\
                \mu_0 H_{\mathrm{ha}}
            \end{pmatrix}
            \cdot
            \begin{pmatrix}
                \cos(\alpha) \\
                \sin(\alpha)
            \end{pmatrix}.
\end{equation}
Using the trigonometric identities $\cos(\pi/2-\epsilon)=\sin(\epsilon)$ and $\sin(\pi/2-\epsilon)=\cos(\epsilon)$ a transformation of coordinates $\alpha=\pi/2-\epsilon$ yields:
\begin{equation}
\label{eq:energy_SW_classical_transf}
   E = Kt\cos( \epsilon)^2 - Mt
            \begin{pmatrix}
                \mu_0 H_{\mathrm{ea}}\\
                \mu_0 H_{\mathrm{ha}}
            \end{pmatrix}
            \cdot
            \begin{pmatrix}
                \sin(\epsilon) \\
                \cos(\epsilon)
            \end{pmatrix}.
\end{equation}

\end{document}